\documentclass[a4paper]{article}

\usepackage{INTERSPEECH2019}
\usepackage{xcolor}
\usepackage{url}
\usepackage{gensymb}
\usepackage{csquotes}
\usepackage{hyperref}
\usepackage{booktabs}
\usepackage{multicol}
\usepackage{multirow}

\title{Ultrasound tongue imaging for diarization and  alignment\\of child speech therapy sessions}
\name{Manuel Sam Ribeiro, Aciel Eshky, Korin Richmond, Steve Renals}
\address{The Centre for Speech Technology Research, University of Edinburgh, UK}
\email{\{sam.ribeiro, aeshky, korin.richmond, s.renals\}@ed.ac.uk}

\begin{document}

\maketitle

\begin{abstract}

We investigate the automatic processing of child speech therapy sessions using ultrasound visual biofeedback, with a specific focus on complementing acoustic features with ultrasound images of the tongue for the tasks of speaker diarization and time-alignment of target words.
For speaker diarization, we propose an ultrasound-based time-domain signal which we call estimated tongue activity.
For word-alignment, we augment an acoustic model with low-dimensional representations of ultrasound images of the tongue, learned by a convolutional neural network.
We conduct our experiments using the Ultrasuite repository of ultrasound and speech recordings for child speech therapy sessions. For both tasks, we observe that systems augmented with ultrasound data outperform corresponding systems using only the audio signal.

\end{abstract}


\section{Introduction}

Developmental speech sound disorders (SSDs) are a common communication impairment in childhood \cite{wren2016prevalence}.
Children with SSDs consistently exhibit difficulties in the production of specific speech sounds in their native language.
Speech disorders have the potential to negatively affect the lives and the development of children.
For example, self-awareness of disordered speech may lead to low-confidence in social situations or introduce communication barriers that lead to increased difficulty in learning and decreased literacy levels \cite{johnson2010twenty, lewis2011literacy, mccormack2011nationally}.

Clinical intervention is available for these children.
Auditory-based techniques are often efficient for pre-schoolers, but for older children with persistent disorders such methods may be unsuccessful \cite{cleland2019enabling}.
For these cases, there is growing evidence that visual biofeedback (VBF) is helpful
\cite{katz2010treating, gibbon2010visual, byun2012investigating, cleland2015using}.
VBF allows the visualization of the vocal tract during the speech production process, enabling patients to correct inaccurate articulations in real-time.
Ultrasound VBF is a clinically safe and non-invasive method that uses standard medical ultrasound to visualize tongue movements \cite{cleland2019enabling, cleland2015using, bernhardt2005ultrasound, furniss2018seeing}.
Besides being useful for patients, ultrasound tongue imaging (UTI) helps therapists in the assessment and diagnosis of SSDs, as it provides information not available in the acoustic signal (e.g., the presence of double articulations or undifferentiated lingual gestures).

The automatic processing of speech therapy data can be helpful to patients and therapists.
Children can use screening tools at an early age to determine whether they need to be assessed by a therapist \cite{sadeghian2015towards, ward2016automated}.
Instrumented methods such as spectrogram or ultrasound analysis can be used to assist the therapist in the assessment, diagnosis, or quantification of treatment efficacy.
However, current practice 
relies on manual annotation by the therapist or other trained professionals.
Examples include the identification of boundaries for target words or phones, using the spectrogram or the ultrasound signal.
This is a laborious process that could be alleviated using automated methods.

There are several challenges associated with the automatic processing of child speech therapy sessions.
They contain dialogue between therapist and child, extraneous child speech, multiple attempts at pronouncing target words, or mispronunciations due to SSDs.
Although there is knowledge of target words, full transcriptions of the recorded speech are not available.
Additionally, there are various other challenges associated with child speech processing \cite{russell2007challenges}, disordered speech processing \cite{christensen2012comparative}, and ultrasound image processing \cite{stone2005guide}.

In this work, we are concerned with the speaker diarization and the time-alignment of target words from child speech therapy sessions using U-VBF.
Robust methods to solve these tasks can alleviate the manual workload of speech therapists, while having the added benefit of preparing data for further processing, such as the development of screening tools.
It is the goal of this paper to investigate methods that \emph{complement the audio signal with ultrasound images of the tongue}.
For diarization, we propose a time-domain signal which we call \emph{estimated tongue activity}.
For word-alignment, we augment an acoustic model with low-dimensional representations of ultrasound tongue images learned by a convolutional neural network.
Section \ref{sec:ultrasuite} introduces our dataset, while Sections \ref{sec:diarization} and \ref{sec:alignment} describe our methods for speaker diarization and word alignment respectively. Section \ref{sec:discussion} provides a discussion and conclusion.

\begin{figure*}[t]
  \centering
  \includegraphics[width=1\linewidth]{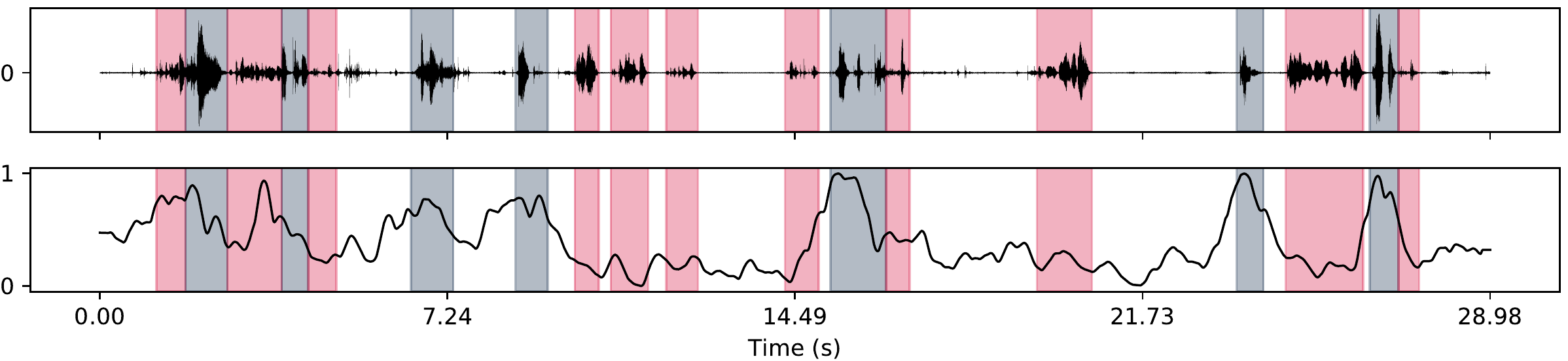}
  \caption{Sample waveform (top) with corresponding unity-based normalized estimated tongue activity (bottom). Highlighted segments denote speech from the child (blue) and the therapist (red).}
  \label{fig:tongue_activity}
\end{figure*}

\section{The UltraSuite repository}
\label{sec:ultrasuite}

UltraSuite\footnote{\url{http://www.ultrax-speech.org/ultrasuite}}\cite{eshky2018ultrasuite} is a repository of synchronized ultrasound and audio data from child speech therapy sessions.
The repository currently contains three data collections.
One includes recordings from 58 typically developing children: Ultrax Typically Developing (UXTD). Additional collections include data from children with speech sound disorders recorded over the course of assessment and therapy sessions: Ultrax Speech Sound Disorders (UXSSD, 8 children) and Ultraphonix (UPX, 20 children).
Each waveform is accompanied by a prompt and corresponding ultrasound recording.
The prompt includes a list of target words to be elicited by the child and does not correspond to a direct transcription of the utterance.
Recordings may include therapist intervention to encourage or guide the patient or extraneous speech by the child (e.g. false starts, multiple attempts at a target word, or unrelated speech). Ultrasound was recorded using an Ultrasonix SonixRP machine using Articulate Assistant Advanced (AAA) software at $\sim$121fps with a 135\degree \ field of view.
A single B-Mode ultrasound frame has 412 echo returns for each of 63 scan lines, giving a 63x412 \enquote{raw} ultrasound frame that captures a midsagittal view of the tongue.

\vspace{-1mm}
\section{Speaker diarization}
\label{sec:diarization}

Speaker diarization aims to identify \textquote{who spoke when} \cite{anguera2012speaker}.
For our scenario, the goal is to identify speech segments belonging to the child or the therapist.
The difficulty of this task  is increased due to very short speech segments with very fast turns between speakers \cite{cristia2018talker}.
In this section, we compare baselines based on voice activity detection (VAD),  diarization methods using i-vectors \cite{sell2014speaker}, and hidden Markov model (HMM) / Gaussian mixture model (GMM) systems bootstrapped by a small set of labeled data \cite{najafian2016speaker}.
One of our key contributions is the inclusion of the ultrasound signal via \emph{estimated tongue activity}.
For evaluation, we use a set of 120 utterances (60 from typically developing and 60 from speech disordered speakers), which have been manually annotated. We measure precision and recall in terms of retrieved child speech and diarization error rate (DER) in terms of silence, child, and therapist speech. Evaluation metrics are computed in terms of time in seconds using a 100ms collar with \emph{pyannote.metrics} \cite{bredin2017pyannote}.

\subsection{Estimated tongue activity (ETA)}
The ultrasound signal visualizes the vocal tract of the child, but it does not capture information regarding other speakers in the room.
We propose a time domain signal which we call \emph{estimated tongue activity} (ETA).
This is computed given a sliding time window over the ultrasound frames.
The variance of each echo return is taken over time, and the overall mean variance of all echo returns is computed.
Higher variance indicates that the content of the ultrasound frames changes rapidly over time.
Figure \ref{fig:tongue_activity} illustrates the ETA signal, considering a sliding window of approximately 160 milliseconds.
Child speech (highlighted in blue) generally occurs with high variance, while therapist speech (in red) occurs with low variance.

\begin{table}[t]
\caption{Results for speaker diarization for typically developing speech (UXTD) and disordered speech (UXSSD) . Precision, recall, and F\textsubscript{1} score are measured in terms of retrieved child speech. Diarization Error Rate (DER) measures  silence and speech, while Confusion Error (Conf) measures child and therapist speech error. Highlighted results indicate best systems.}
\label{tab:diarization-results}
\centering
\resizebox{0.96\columnwidth}{!}{%
\begin{tabular}{@{}lccc|cc@{}}
\toprule
\textbf{Systems}                               & \textbf{Prec} & \textbf{Rec} &  $\mathbf{F_1}$  & \textbf{DER} & \textbf{Conf} \\ \midrule
\multicolumn{6}{c}{\textbf{Typically developing speech}}                                                                                                                                                                           \\ \midrule
VAD                                            & 0.89                              & 0.93                             & 0.91                            & 18.7                            & 1.5                             \\
VAD+ETA                                        & 0.92                              & 0.66                             & 0.77                            & 37.1                            & 19.9                            \\
PLDA i-vectors (mfcc)                                  & 0.89                              & 0.80                             & 0.85                            & 23.3                            & 6.2                             \\
\quad+ $f_0$                                  &  0.89                              & 0.70                             & 0.78                            & 26.8                            & 9.8                             \\
HMM-GMM (mfcc)                                 & 0.86                              & 0.72                             & 0.79                            & 38.0                            & 0.8                             \\
\quad+ $f_0$                                  & 0.83                              & 0.81                             & 0.87                            & 26.0                            & 0.2                             \\
\quad+ $f_0$ + ETA                            & 0.93                              & 0.90                             & \textbf{0.92}                            & \textbf{17.1}                            & 0.4                             \\
\quad+ $f_0$ + ETA + semi-sup                 & 0.98                              & 0.84                             & 0.90                            & 18.5                            & \textbf{0.0}                             \\ \midrule
\multicolumn{6}{c}{\textbf{Disordered speech}}                                                                                                                                                                               \\ \midrule
VAD                                            & 0.55                              & 0.90                             & 0.68                            & 47.6                            & 29.7                            \\
VAD+ETA                                        & 0.81                              & 0.76                             & 0.79                            & 32.0                            & 14.1                            \\
PLDA i-vectors (mfcc)                                  & 0.67                              & 0.71                             & 0.69                            & 32.4                            & 14.6                            \\
\quad+ $f_0$                                  &  0.69                              & 0.75                             & 0.72                            & 34.4                            & 16.6                            \\
HMM-GMM (mfcc)                                 & 0.57                              & 0.71                             & 0.63                            & 52.5                            & 12.5                            \\
\quad+ $f_0$                                  & 0.82                              & 0.81                             & 0.82                            & 29.1                            & 8.6                             \\
\quad+ $f_0$ + ETA                            & 0.81                              & 0.89                             & 0.85                            & \textbf{24.0}                            & 8.4                             \\
\quad+ $f_0$ + ETA + semi-sup                 & 0.95                              & 0.81                            & \textbf{0.87}                           & 28.2                            & \textbf{0.2}                             \\ \bottomrule
\end{tabular}
}
\end{table}

\subsection{Experiments and results}
\textbf{Baselines.} 
We consider threshold-based baselines using frame-level energy-based voice activity detection (VAD), available in Kaldi \cite{povey2011kaldi}, and the proposed estimated tongue activity  (ETA).
In these systems, we use a threshold of 7 to perform VAD and a threshold of 0.5 to perform tongue activity detection using ETA.
For the method using VAD, we simply term all activity above the threshold as child speech.
For the method combining both signals (\emph{VAD + ETA}), all frames below the VAD threshold are taken as silence and frames that are above the threshold are further processed using the ETA signal. Those frames above the ETA threshold are identified as child speech, while those below are assumed to be therapist speech.

\smallskip\noindent
\textbf{i-vectors.} This baseline follows the method presented by Sell and Garcia-Romero~\cite{sell2014speaker}, available in Kaldi \cite{povey2011kaldi}.
We use the PF-STAR children's speech corpus \cite{batliner2005pf_star} (7.4 hours, 86 speakers) to train a Universal Background Model (GMM with 512 components) and extract 64 dimensional i-vectors over 1.5 second windows.
We use in-domain heldout data from UltraSuite to normalize the training data for the probabilistic linear discriminant analysis (PLDA) scoring and perform agglomerative clustering assuming 2 speakers in each recorded session.

\smallskip\noindent
\textbf{HMMs.} Our proposed HMM approaches require less data than typical i-vector clustering approaches. We use the transcriptions available with the UXTD dataset, which contain tags denoting whether a word is spoken by the therapist or the child \cite{eshky2018ultrasuite}.
Each word is reduced to \enquote{child} or \enquote{therapist} tokens, corresponding to speaker turns in each utterance. 
These are modelled with 5 state ergodic HMMs with 1000 Gaussian components.
We allow HMMs for child, therapist, and optional silence and noise.
Turn-taking sequences are derived from a transcription and there is no initial alignment with respect to the features.
We evaluate this method using a combination of 20 MFCCs, 3 $f_0$ features \cite{ghahremani2014pitch}, and the ETA signal.
Because we bootstrap HMM/GMMs with a small set of annotated data (1.75 hours), we further investigate a \emph{semi-supervised} approach incorporating additional unlabeled disordered speech data, in which a model trained on typically development data is used to decode the UPX dataset (11.05 hours), with the hypothesized labels then used to retrain the HMM/GMM.
For all systems, a post-processing step merges identical labels separated by silence shorter than 100ms and removes labels with duration less than 50ms.

\begin{figure}[t]
  \centering
  \includegraphics[width=0.98\columnwidth]{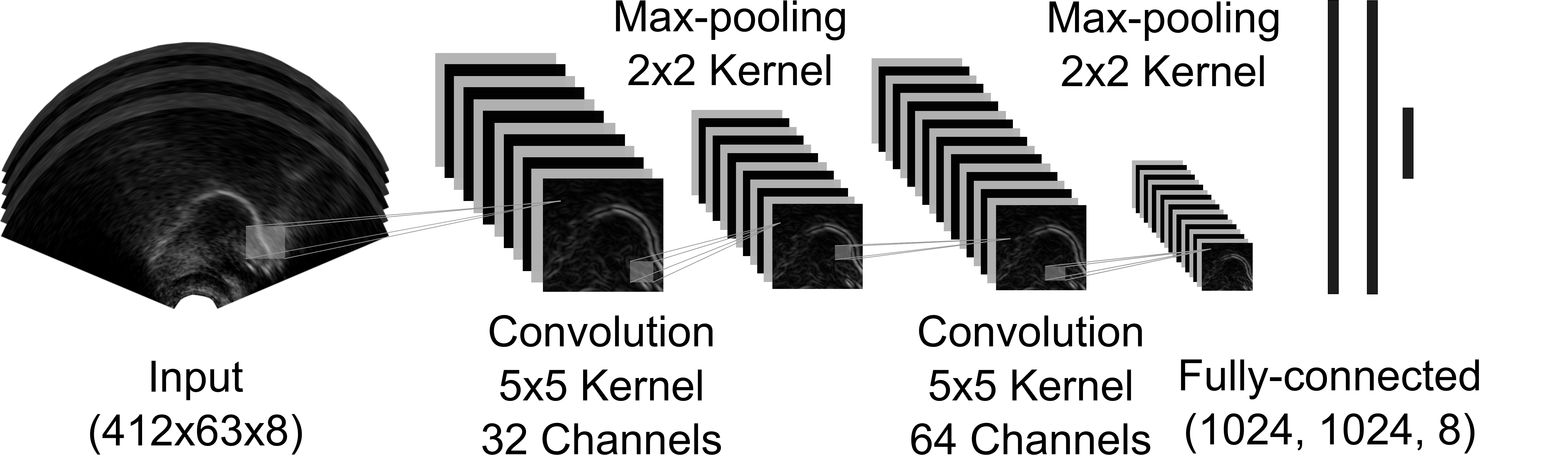}
  \caption{Convolutional neural network architecture for ultrasound embedding extraction.}
  \label{fig:cnn-arch}
\end{figure}

\smallskip
\noindent
\textbf{Results} are presented in Table \ref{tab:diarization-results}.
The simple baseline approaches perform well for the typically developing dataset, which has minimal adult intervention, and therefore less need for diarization \cite{eshky2018ultrasuite}.
The loss in recall with ETA thresholding is caused by losing segments with reduced tongue activity (e.g. words such as \enquote{apa}).
In the case of the disordered speech dataset, therapist intervention is more frequent, and ETA is a useful addition to VAD.
The i-vector system does not perform as well as expected.
This might be due to size of the database used to train the UBM, but also to the challenges of fast speaker turns. 
Similar results were found using a large adult database (NIST SRE datasets, similar to \cite{sell2014speaker}).
We hypothesize that the i-vectors are not robust enough, as we enforce a very small window due to the rapid turns between speakers.
High error rates have previously been found for a similar scenario \cite{cristia2018talker}.
Allowing a longer window would lead to equally poor i-vectors, as they would be computed over a mixture of adult and child speech.
Surprisingly, including $f_0$ was not helpful using this method.

The proposed HMM-GMM systems achieved better results, especially when using additional features.
The use of $f_0$ leads to  good improvements, which are complemented by the inclusion of the ETA signal.
Overall, a basic HMM-GMM using a combination of all features achieves good performance on both datasets.
The semi-supervised approach leads to very low speaker confusion, but DER is increased due to missed cases.
Although the model becomes stronger at differentiating speakers, it does so at the expense of the silence and noise models.

\vspace{-1mm}
\section{Word alignment}
\label{sec:alignment}

Word alignment involves time-aligning target words, given in the prompt, with the speech signal.
There are various challenges associated with this task when applied to speech therapy sessions.
This is noticeable when force-aligning the audio with the expected prompt, using baseline standard methods.
In this section, we complement the audio signal with ultrasound images of the tongue for acoustic modeling.

\subsection{Ultrasound embeddings}
\label{sec:embeddings}

We augment the audio features with ultrasound embeddings extracted from a convolutional neural network (CNN).
The network uses the UXTD dataset as training data and follows the architecture and findings reported by \cite{ribeiro2019speaker}, illustrated in Figure \ref{fig:cnn-arch}.
Input consists of 8 ultrasound frames, grouped as multiple channels. These are: the current frame, 3 left and right neighboring frames, and the speaker mean.
Two sets of convolution and max-pooling layers are followed by three fully-connected layers.
The final layer of the network is 8 dimensional, and we let this representation be the ultrasound embedding for the current timestamp.
Output classes roughly correspond to place of articulation, defined over the entire phoneset.

\begin{table}[t]
\caption{Results for word alignment (precision, recall, $F_1$) and recognition (WER) for typically developing speech (UXTD) and disordered speech (UXSSD).  Baseline systems use GMMs trained on Ultrasuite and PF-Star data.  USuite systems use neural network acoustic models.
Highlighted results indicate best performing systems.
}
\label{tab:alignment-results}
\centering
\resizebox{0.93\columnwidth}{!}{%
\begin{tabular}{@{}lccc|c@{}}
\toprule
\textbf{Systems} & \textbf{Pre} & \textbf{Rec} & $\mathbf{F_1}$ & \textbf{WER}                 \\ \midrule
\multicolumn{5}{c}{\textbf{Typically developing speech}}                                                    \\ \midrule
Baseline (no diarization)               & 0.88         & 0.90         & 0.89             & 36.29    \\
Baseline                                & 0.89         & 0.91         & \textbf{0.90}    & 37.18        \\
USuite                                  & 0.86         & 0.92         & 0.89         & 31.35                 \\
USuite+PFstar                           & 0.85         & 0.92         & 0.89         & 30.96                 \\
USuite+UTI                              & 0.86         & 0.92         & \textbf{0.90}         & 31.22                 \\
System combination ($\alpha=0.6$)       & 0.86         & 0.93         & 0.89         & \textbf{28.68}        \\
System combination ($\alpha=0.7$)       & 0.86         & 0.93         & 0.89         & 29.44     \\ \midrule
\multicolumn{5}{c}{\textbf{Disordered speech}}                                                                                       \\ \midrule
Baseline (no diarization)                     & 0.53        & 0.46        & 0.50       & 71.19     \\
Baseline                & 0.77        & 0.68        & 0.72       & 69.24     \\
USuite                  & 0.75        & 0.70        & 0.73       & 63.21     \\
USuite+PFstar           & 0.76        & 0.72        & 0.74       & 59.35     \\
USuite+UTI              & 0.75        & 0.70        & 0.72       & 62.73     \\
System combination ($\alpha=0.6$)     & 0.77        & 0.73       & \textbf{0.75}       & 59.35     \\
System combination ($\alpha=0.7$)     & 0.77        & 0.72       & \textbf{0.75}       & \textbf{58.14}     \\ \bottomrule
\end{tabular}
}
\end{table}

\subsection{Acoustic model training and evaluation}

We train the acoustic models with the Kaldi speech recognition toolkit \cite{povey2011kaldi}.
GMMs are initialized on pooled data from UXTD (1.75 hours), UPX (11.05 hours), and the training set of the PF-STAR children's speech corpus (7.4 hours) \cite{batliner2005pf_star}.
The inclusion of the PF-STAR corpus is beneficial for model initialization as it contains additional manually transcribed data for 86 speakers.\footnote{We have attempted to initialize models using only in-domain data, but including the PF-STAR corpus lead to a stronger initialization at this stage. We therefore opt to treat that as our baseline and we conduct our analyses over the neural network acoustic models.}
For UXTD and UPX data, we remove speech frames associated with the therapist using the system with the best DER from Section \ref{sec:diarization}.
This way, we avoid aligning target words with time segments belonging to the speech therapist.
After monophone and triphone training, Mel Frequency Cepstral Coefficients (MFCCs) are processed with Linear Discriminant Analysis (LDA) and a Maximum Likelihood Linear Transform (MLLT).
This is followed by Speaker Adaptive Training (SAT) with feature-space MLLR (fMLLR) \cite{rath2013improved, fainberg2016improving}.
This HMM-GMM system is denoted \textbf{Baseline} in Table \ref{tab:alignment-results}.
Alignment and fMLLR features from this system are then used to train feedforward neural networks, following Kaldi's \emph{nnet1} recipe.

For evaluation, we use a set of 398 utterances (199 from typically developing speakers, and 199 from speakers with speech disorders), manually annotated with word boundaries from target words given by the prompt.
Precision and recall are measured in terms of correctly retrieved time with respect to the manual gold standard, given matching word labels \cite{bredin2017pyannote}.
Additionally, in order to further quantify the robustness of the acoustic models, we consider an experimental scenario where we decode within oracle word boundaries.
For this case, we use a uniform unigram language model, built from the joint vocabulary of the UXTD/UXSSD datasets (1161 words).
Results are reported in terms of word error rate (WER).

\begin{figure}[t]
  \centering
  \includegraphics[width=0.95\columnwidth]{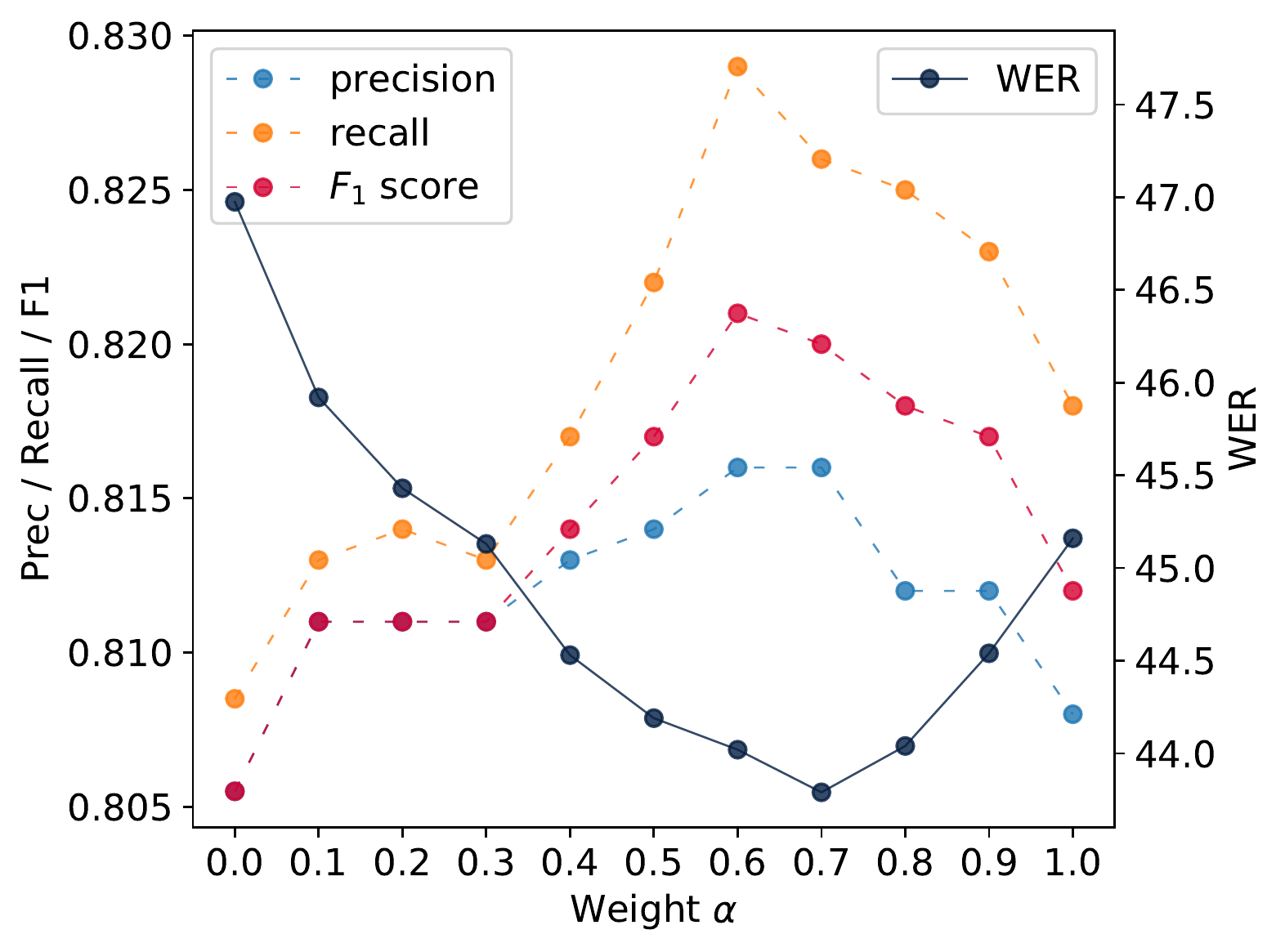}
  \caption{Combination of models augmented with out-of-domain acoustic data (USuite+PFstar) with weight $\alpha$ and augmented with ultrasound embeddings (USuite+UTI) with weight $(1-\alpha)$. Results are averaged across typically developing and speech disordered evaluation sets.}
  \label{fig:model-combination}
\end{figure}

\subsection{Experiments and results}
We compare various neural network acoustic models.
The first uses only acoustic in-domain data from speech therapy sessions (UXTD and UPX training data), termed \textbf{USuite}.
This training set is also augmented with out-of-domain acoustic data from the PF-STAR children's speech corpus (\textbf{USuite+PFstar}).
We augment the in-domain system with the ultrasound embeddings described in Section \ref{sec:embeddings} using a context of 4 frames. Acoustic and ultrasound features are concatenated for the neural network (\textbf{USuite+UTI}).
Ideally we would include all data sources, but ultrasound does not exist in the PF-STAR corpus. 
Therefore, we combine the two systems by interpolating their posterior probabilities, with weight $\alpha$ for USuite+PFstar and $(1-\alpha)$ for USuite+UTI (\textbf{System combination}).

Results are presented in Table \ref{tab:alignment-results}.
Diarization is not beneficial for the typically developing dataset, as it has a small amount of adult speech. 
This is not the case for disordered speech, where diarization leads to more accurate alignment and recognition.
Although baseline GMMs perform reasonably well in terms of alignment, they do not outperform neural network acoustic models in terms of recognition.
Augmenting in-domain data with ultrasound leads to improvements on both datasets, but not as much as using additional out-of-domain acoustic data.
The best system is a combination of the systems augmented with out-of-domain data and ultrasound embeddings.
Although we observe small improvements in terms of alignment metrics, WER shows that combined models are more robust than either of them separately.
Figure \ref{fig:model-combination} shows results for model combination across the range of $\alpha$ values.
Best results occur when there is a slight bias towards the system augmented with out-of-domain training data.
These results also show that ultrasound and audio data can complement each other well.

\begin{table}[t]
\caption{Results for combination of models with $\alpha=0.7$ for disordered speech (UXSSD), grouped by assessment session at various stages of therapy. Post-therapy is recorded immediately after therapy and Maintenance several months after.}
\label{tab:ssd-results}
\centering
\resizebox{0.70\columnwidth}{!}{%
\begin{tabular}{lccc|c}
\toprule
\textbf{Sessions}   & \textbf{Pre} & \textbf{Rec} &  $\mathbf{F_1}$  & \textbf{WER}  \\ \midrule
Baseline            & 0.77        & 0.72        & 0.75       & 62.01         \\
Mid-therapy         & 0.82        & 0.76        & 0.80       & 60.90         \\
Post-therapy        & 0.80        & 0.74        & 0.77       & 47.89         \\
Maintenance         & 0.71        & 0.69        & 0.70       & 60.35         \\ \midrule
Global              & 0.77        & 0.72        & 0.75       & 58.14         \\ \bottomrule
\end{tabular}
}
\end{table}


\vspace{-1mm}
\section{Discussion and conclusions}
\label{sec:discussion}

Speech therapy using U-VBF is beneficial for patients, who may visualize their articulators during speech production, and therapists, who may use ultrasound for more accurate diagnoses and treatments.
The automatic processing of speech therapy sessions provides additional benefits by alleviating some of the time-consuming manual tasks undertaken by therapists, as well as preparing data for further processing.
In this paper we proposed methods that use ultrasound tongue imaging to complement the audio signal for speaker diarization and the time-alignment of target words.
For both tasks, we have observed improvements in models augmented with ultrasound when compared with models using only audio recordings.

For speaker diarization, we have proposed a measure of Estimated Tongue Activity (ETA) which we have used for tongue activity detection.
In the future, this signal could be useful to measure patients' overall tongue activity during therapy sessions.
For example, to investigate whether children shadow therapists or whether they exhibit a larger amount of tongue movement after therapy.
One of the main issues for diarization, however, is the lack of annotated training data.
These methods are designed to be used by therapists during and after assessment sessions, and they could learn from user input.
For example, a single session typically contains the same two speakers.
Adaptation based on a small amount of data annotated by the therapist could help with speaker diarization.

In terms of alignment, we showed that a combination of models augmented with out-of-domain acoustic data and in-domain ultrasound data led to more robust acoustic models.
Table \ref{tab:ssd-results} shows results by assessment session at various stages of therapy.
These results should not be directly compared, as they contain different target words, but they do illustrate the temporal dependency of therapy sessions, which could be used for longitudinal online learning \cite{karanasou2015speaker}.
Additionally, there are various issues that were not addressed by this paper.
Future work may consider learning speaker- or session-specific pronunciations \cite{christensen2013learning}, as
baseline sessions are expected to have different pronunciations than post-therapy or maintenance sessions.
Furthermore, insertions and deletions with respect to the prompt could be handled by relaxing the constraints of the linear transducer used for alignment \cite{bell2015system}.

\vspace{-1mm}
\section{Acknowledgements}
Supported by the EPSRC Healthcare Partnerships Programme grant number EP/P02338X/1 (Ultrax2020).

\vfill\pagebreak
\bibliographystyle{IEEEtran}
\bibliography{references}

\end{document}